\input{aipcheck}

\documentclass[
    ,final            
  ]
  {aipproc}
\usepackage{amsmath,amssymb,verbatim,epsfig,psfig,graphicx,rotating}
\newcommand{\be}{\begin{equation}}
\newcommand{\ee}{\end{equation}}
\layoutstyle{6x9}

\begin{document}

\title{Primordial Magnetic Fields from Out of Equilibrium Cosmological Phase
Transitions\footnote{Lecture given at the International Conference 
Magnetic Fields in the Universe, Angra dos Reis, Brazil, November, 2004}}


\author{D. Boyanovsky}{
  address={Department of Physics and
Astronomy, University of Pittsburgh, Pittsburgh, \\ Pennsylvania
15260, USA. \\ Observatoire de Paris, LERMA. Laboratoire
Associ\'e au CNRS UMR 8112.
 \\61, Avenue de l'Observatoire, 75014 Paris, France.}
}

\author{\underline{H. J. de Vega}}{
  address={LPTHE, Universit\'e
Pierre et Marie Curie (Paris VI) et Denis Diderot (Paris VII),
Laboratoire Associ\'e au CNRS UMR 7589,
Tour 24, 5\`eme. \'etage, 4, Place Jussieu, 75252 Paris, Cedex 05,
France.}, altaddress={Department of Physics and
Astronomy, University of Pittsburgh, Pittsburgh, \\ Pennsylvania
15260, USA. \\ Observatoire de Paris, LERMA. Laboratoire
Associ\'e au CNRS UMR 8112.
 \\61, Avenue de l'Observatoire, 75014 Paris, France.}
}
\begin{abstract}
The universe cools down monotonically following its expansion.
This generates a sequence of phase transitions. If a second order phase transition
happens during the radiation dominated era with a charged order parameter,
spinodal unstabilities generate large numbers of charged
particles. These particles hence produce magnetic fields.
We use out of equilibrium field theory methods to study the dynamics
in a mean field or large $N$ setup. The
dynamics after the transition features two distinct stages: a
spinodal regime dominated by linear long-wavelength instabilities,
and a scaling stage in which the non-linearities and backreaction
of the scalar fields are dominant. This second stage describes the
growth of horizon sized domains. We implement a formulation based
on the non-equilibrium Schwinger-Dyson equations to obtain the
spectrum of magnetic fields that includes the dissipative effects
of the plasma. We find that large scale magnetogenesis is
efficient during the scaling regime. Charged scalar field
fluctuations with wavelengths of the order of the Hubble radius
induce large scale magnetogenesis via \emph{loop effects}. The
leading processes are: pair production, pair annihilation
and low energy bremsstrahlung, these processes while forbidden in
equilibrium are allowed strongly out of equilibrium. The ratio
between the energy density on scales larger than $L$ and that in
the background radiation $r(L,T)= \rho_B(L,T)/\rho_{cmb}(T)$ is
$r(L,T) \sim 10^{-34}$ at the Electroweak scale and $r(L,T)\sim
10^{-14}$ at the QCD scale for $L \sim 1~\mbox{Mpc}$. The
resulting spectrum is insensitive to the magnetic diffusion length
and equipartition between electric and magnetic fields does
{\bf not} hold. We conjecture that a similar mechanism could be
operative after the QCD chiral phase transition.
\end{abstract}

\maketitle


\section{Early Cosmology and Fundamental Physics}

The history of the universe is determined by its expansion and
consequent cooling. During most of its early history the Universe
 is homogeneous and isotropic to an excellent approximation and
is therefore described by the spatially flat
Friedmann-Robertson-Walker (FRW) geometry
\begin{equation}\label{FRW}
ds^2= dt^2-a^2(t) \; d{\vec x}^2
\end{equation}
where the scale factor $ a(t) $ grows with $t$. Physical lengths are
proportional to $  a(t) $ and the temperature decreases as $ T(t)
\sim \frac{1}{a(t)} $.  This monotonous decrease of the temperature
generates a sequence of cosmological {\bf phase transitions} with
the ensuing breaking of internal symmetries. The symmetry of the
Universe reduces through each phase transition.

The main ingredients  to describe the early Universe are:

\begin{itemize}
\item{General Relativity: Einstein's Theory of Gravity}

The matter distribution determines the geometry of the spacetime
through the Einstein equations. For the geometry eq.(\ref{FRW}), the
Einstein equations reduce to one scalar equation, the
Einstein-Friedman equation \be \label{ef} \left[ \frac{1}{a(t)} \;
\frac{da}{dt} \right]^2 = \frac{8 \, \pi}{3} \; G \; \rho(t) \; ,
\ee where $G$ stands for Newton's gravitational constant and $
\rho(t) $ for the energy density.

\item{Quantum Field Theory and String Theory to describe Matter}

Since the energy scale in the early universe is so high (well beyond
the rest mass of particles), a quantum field theoretical description
for matter is unavoidable. Only such context permits a correct
description of particle production and  particle decays.

\end{itemize}

\bigskip

Electromagnetic, weak and strong interactions are well described by
the so-called standard model. That is, quantum chromodynamics (QCD)
combined with the electroweak theory (electromagnetic and weak
interactions). This a non-abelian gauge theory associated to the
symmetry group $ SU(3) \otimes SU(2) \otimes U(1) $. The $SU(3)$
corresponding to the color group of QCD while $ SU(2) \otimes U(1) $
describes the electroweak sector. To this scheme, one adds presently
neutrino masses (through the see-saw mechanism) to explain  neutrino
oscillations.

The energy scale in QCD is about $ \sim 100$MeV $\simeq 10^{12}$K
corresponding to the chiral symmetry breaking and determined by the
pion mass, while the energy scale for the electroweak is the Fermi
scale $\sim 100$GeV $\simeq 10^{15}$K, which is determined by the
mass of the vector bosons.

The standard model has been verified experimentally with spectacular
precision. However, it is an incomplete quantum field theory and it
is the major challenge of our times to understand its extension.  It
seems obvious that extensions of the Standard model will be
symmetric under a group containing $ SU(3) \otimes SU(2) \otimes
U(1) $ as a subgroup. Proposals for a Grand Unified Theory (GUT)
include $ SO(10), \; SU(6) $ and $E_6$ as symmetry group.

The grand unification idea consists in that at some energy scale all
three couplings (electromagnetic, weak and strong) should become of
the same strength.  The running of the couplings with the energy (or
the length) is governed
 by the renormalization group.
For the standard model of electromagnetic, weak and strong
interactions, the renormalization group yields that the three
couplings get unified approximately at $ \sim 10^{16}$GeV \cite{sw}.
A better convergence is obtained in supersymmetric extensions of the
standard model\cite{sw}.

Grand unified models may posses magnetic monopoles or {\bf not}
according to their symmetry group and to the symmetry of the ground
state. Notice that no experimental evidence for magnetic monopoles
has been found so far.

Quite generally, the internal  symmetry increases with energy. This
is true in general, in statistical mechanics, condensed matter as
well as in cosmology. For example, a ferromagnet at temperatures
higher than the Curie point is in the symmetric phase with zero
magnetization. Below the Curie point, the internal symmetry is
spontaneously broken by a non-zero spontaneous magnetization.

 Current models purport that the universe started with maximal symmetry before inflation and this
symmetry reduces gradually while the universe expands and cools. The
symmetry breaking transitions includes both the {\bf internal
symmetry groups} (as the GUT's symmetry group that eventually
reduces to the $ SU(3) \otimes SU(2) \otimes U(1) $ group) as well
as the translational and rotational symmetries which are broken by
the density fluctuations amplified by gravitational instabilities
leading to  structure formation.

It should be noticed, however, that no direct manifestation of
supersymmetry is known so far. An indication emerges by studying the
energy running of the (electromagnetic, weak and strong) in the
standard model and in its minimal supersymmetric extension (MSSM).
All three couplings meet at $ E \simeq  2 \times 10^{16}$GeV in the
MSSM. The coupling unification becomes quite loose in the Standard
Model. This is why the renormalization group running of the
couplings in the MSSM supports the idea that supersymmetry would be
a necessary ingredient of a GUT.

Neutrino oscillations and neutrino masses are currently explained in
the see-saw mechanism as follows\cite{ita},
$$
\Delta m_{\nu} \sim \frac{M^2_{Fermi}}{M}
$$
where $ M_{Fermi} \sim 250$ GeV is the Fermi mass scale, $M \gg
M_{Fermi} $ is a large energy scale and $ \Delta m_{\nu} $ is the
difference between the neutrino masses for different flavors. The
observed values for $ \Delta m_{\nu} \sim 0.009 - 0.05 $ eV
naturally call for a mass scale $ M \sim 10^{15-16}\;  \textit{GeV}$
close to the GUT scale\cite{ita}.

The third evidence for an energy scale about $ 10^{16}$GeV comes
from inflation: data on the cosmic microwave background (CMB)
anisotropies indicate that the inflation scale, the grand unification scale and the
supersymmetry breaking scale actually coincide \cite{nos}.

\section{Primordial Seeds for the Magnetic Fields in the Universe}

A variety of astrophysical observations including Zeeman splitting,
synchrotron emission, Faraday rotation measurements (RM) combined
with pulsar dispersion measurements (DM) and polarization
measurements suggest the presence of large scale magnetic
fields\cite{cmp,han}. The  strength of typical galactic magnetic
fields is measured to be $\sim \mu~G$\cite{cmp,han} and they are
correlated on very large scales up to galactic or even larger
reaching to scales of cluster of galaxies $\sim
1~\mbox{Mpc}$\cite{cmp}. The origin of these large scale magnetic
fields is still a subject of much discussion and controversy. It is
currently agreed that a variety of dynamo mechanisms are efficient
in {\bf amplifying} seed magnetic fields with typical growth rates
$\Gamma \sim \mbox{Gyr}^{-1}$ over time scales $\sim 10-12 $ Gyr
(for a thorough discussion of the mechanisms and  models
see\cite{cmp}). The ratio of the energy density of the seed magnetic
fields  on scales larger than $L$ (today) to that in the cosmic
background radiation, $r(L)=\rho_B(L)/\rho_{cmb}$ must be $r(L\sim
1\mbox{Mpc})\geq 10^{-34}$ for a dynamo mechanism to amplify it to
the observed value, or $r(L\sim 1\mbox{Mpc})\geq 10^{-8}$ for the
seed to be amplified solely by the gravitational collapse of a
protogalaxy\cite{cmp}.

There are also different  proposals to explain the origin of the
initial seed. Astrophysical batteries rely on gradients of the
charge density concentration and pressure and their efficiency in
producing seeds of the necessary amplitude is still very much
discussed\cite{cmp}. Primordial magnetic fields that could be the
seeds for dynamo amplification can be generated at different stages
in the history of the early Universe, in particular during
inflation, preheating and or phase transitions\cite{cmp}. Primordial
(hyper) magnetic fields may have important consequences in
electroweak baryogenesis\cite{grassoEW}, Big Bang nucleosynthesis
(see\cite{cmp}), the polarization of the CMB\cite{durrer1} via the
same physical processes as Faraday rotation, and structure
formation\cite{cmp,dolgov2}, thus sparking an intense program to
study the origin and consequences of the generation of magnetic
fields in the early Universe\cite{hogan}-\cite{ahonen}.

A reliable estimate of the amplitude and correlations of seed
magnetic fields must include the dissipative properties of the
plasma, in particular the conductivity\cite{turnerwidrow,Giovanni}.
In ref.\cite{magfiI} we have  introduced a formulation that allows
to compute the generation of magnetic fields from processes strongly
out of equilibrium. This formulation, which is based on the exact
set of Schwinger-Dyson equations for the transverse photon
propagator is manifestly gauge invariant and is general for any
matter fields and any cosmological background (conformally related
to Minkowski space-time). In the case in which strongly out of
equilibrium effects arise from long-wavelength fluctuations, such as
during phase transitions, this formulation allows to separate the
contribution of the hard degrees of freedom which are in local
thermodynamic equilibrium from that of the soft degrees of freedom
that fall out of LTE (local thermal equilibrium) during the phase
transition and whose dynamics is strongly out of equilibrium. This
separation of degrees of freedom leads to a consistent incorporation
of the dissipative effects via the conductivity (for details
see\cite{magfiI}). In that reference a  study of magnetogenesis in
Minkowski space-time during a supercooled phase transitions was
presented and the results highlighted the main aspects of the
generation of magnetic and electric fields in these situations.

We study the generation of large scale (hyper) magnetic fields by a
cosmological phase transition during a radiation dominated era by
implementing the formulation introduced in ref.\cite{magfiI}. The
setting is a theory of $N$ charged scalar fields coupled to an
abelian gauge field (hypercharge). We consider the situation when
this theory undergoes a phase transition after the reheating stage
and before either the Electroweak or the QCD phase transition, since
we expect that these transitions will lead to new physical
phenomena. The non-perturbative dynamics out of equilibrium is
studied in the limit of a large number $N$ of (hyper) charged fields
and to leading order in the gauge coupling. The non-equilibrium
dynamics of the charged scalar sector features two distinct stages.
The first one describes the early and intermediate time regime and
is dominated by the spinodal instabilities which are the hallmark of
the process of phase separation and domain formation and growth.
This stage describes the dynamics between the time at which the
phase transition takes place and that at which non-linearities
become important via the backreaction. The second stage corresponds
to a \emph{scaling regime} which describes the slower
non-equilibrium evolution of Goldstone bosons and the process of
phase ordering\cite{scaling} and growth of horizon-sized domains.
This scaling regime is akin to the solution found in the
\emph{classical} evolution of scalar field models with broken
continuous symmetries after the phase transition that form the basis
for models of structure formation based on topological
defects\cite{turok,durrer}.

The solution of the scalar field dynamics \cite{scaling} is the
input in the expression for the spectrum of the magnetic field
obtained in~\cite{magfiI} to obtain the amplitude of the primordial
seed generated during both stages.

We find that scaling stage is the most important  for the generation
of large scale magnetic fields. Large scale magnetic fields are
generated via loop effects from the dynamics of modes that are at
the scale of the horizon or smaller. The leading order processes
that result in the generation of large scale magnetic fields are: i)
pair {\bf production}, ii) pair annihilation and iii) low energy
bremsstrahlung. These processes  would be forbidden in equilibrium
by energy momentum conservation, but they are allowed strongly out
of equilibrium because of the rapid time evolution of the
cosmological background and the fast dynamics of the scalar field
fluctuations.

The resulting spectrum is rather insensitive to the diffusion length
scale which is much smaller than the horizon during the radiation
dominated era. The ratio of the magnetic energy density on scales
larger than $L$ (today) to the energy density in the background
radiation $r(L,\eta)=\rho_B(L,\eta)/\rho_{cmb}(\eta)$ is summarized
in a compact formula [eq.(\ref{rreges})]. For $L \sim 1~\mbox{Mpc}$
(today) we find $r(L,\eta) \sim 10^{-34}$ at the Electroweak scale
and $r(L,\eta)\sim 10^{-14} $  at the QCD scale, suggesting the
possibility that these primordial seeds could be amplified by dynamo
mechanisms to the values of the magnetic fields consistent with the
observed ones on these scales.

\section{The physical picture}

The extreme energy scale and energy density during the inflationary
and radiation dominated eras call for a quantum field theoretic
treatment of the matter and radiation while the geometry is
described by the classical metric eq.(\ref{FRW}). The fast expansion
of the universe can lead to out of thermal equilibrium situations,
which require the implementation  of out of equilibrium methods. In
addition, nonperturbative methods are needed since the energy
density is proportional to the inverse of the coupling. We developed
nonperturbative field theory methods that successfully treats the
inflationary era in various relevant scenarios and allowed to
compute the primordial perturbations and the CMB fluctuations as
well as to make contact with the customary slow-roll classical
treatment \cite{cosmo}.

We consider here a scalar field carrying an abelian charge and coupled to
the electromagnetic field in the radiation dominated era. 
During the second order phase transition
the concavity of the potential becomes negative at the origin and
strong spinodal fluctuations are generated. These fluctuations in
turn generate a magnetic field with a typical wavelength
corresponding to the wavelength of the spinodally unstable modes.

This  is the main premise of our work\cite{magfiI}: the
instabilities which are the hallmark of a non-equilibrium symmetry
breaking phase transition lead to strong fluctuations of the charged
scalar fields which in turn,  lead to the generation of magnetic
fields through the non-equilibrium evolution.

The main ingredients developed in ref. \cite{magfiI} to compute the
generation of magnetic fields through this non-equilibrium process
were:

\begin{itemize}
\item{A consistent framework to compute the spectrum of generated
magnetic field, namely $\langle \vec{B}({\vec k},t)\cdot
\vec{B}({-\vec k},t)\rangle/V$ with $\vec{B}({\vec k},t)$ the
spatial Fourier transform of the Heisenberg magnetic field
\emph{operator} and $V$ the (comoving) volume of the system.}
\item{Plasma effects were included to assess
the generation and eventual decay of the magnetic fields. For a
large conductivity in the medium, the magnetic field diffuses but
also its generation is hindered. This point is of particular
importance within the cosmological
setting\cite{cmp,Giovanni,turnerwidrow}. }

\item{A major challenge of \emph{any} mechanism of large scale
magnetogenesis is to generate the seed magnetic fields from
microscopic, \emph{causal} processes. An important aspect of the
results presented here is that this generation mechanism is mediated
by \emph{loop} effects and correspond to processes that are
forbidden in equilibrium but allowed strongly out of equilibrium.  }
\end{itemize}

\section{Field Theoretical Model for
Magnetic fields in Friedmann-Robertson-Walker cosmology}

We will not attempt to study a particular gauge theory
phenomenologically motivated by some GUT scenario, but will focus
our study on a generic scalar field model in which the scalar fields
carry an abelian charge. The simplest realization of such model is
scalar electrodynamics with $N$ charged scalar fields $ \phi_r, \;
r=1, \ldots, N$ and one neutral scalar field $ \psi $ whose
expectation value is the order parameter associated with the phase
transition. The neutral field is not coupled to the gauge field and
its acquiring an expectation value does not break the $U(1)$ gauged
symmetry. This guarantees that the abelian gauge symmetry identified
with either hypercharger or electromagnetism is \emph{not
spontaneously broken} to describe the correct low energy sector with
unbroken $U(1)_{EM}$. We will take the neutral and the $N$ complex
(charged) fields to form a scalar multiplet under an $O(2N+1)$
isospin symmetry. The electromagnetic coupling explicitly breaks the
$O(2N+1)$ symmetry down to $SU(N)\times U(1)$. In the absence of
electromagnetic coupling as the neutral field acquires an
expectation value the isospin symmetry is spontaneously broken to
$O(2N)$.  Since by construction only the neutral field acquires a
non vanishing expectation value under the isospin symmetry breaking
the photon remains massless (it will obtain a Debye screening mass
from medium effects.

The action that describes this theory in a general cosmological
background and using conformally rescaled fields is given by
\begin{eqnarray}\label{confoS}
&&S= \int d\eta\; d^3x \left[ \frac{1}{2}\partial_{\mu}\Psi
\partial^{\mu}\Psi+
D_{\mu}\Phi^*D^{\mu}\Phi-M^2(\eta)\left(\frac{\Psi^2}{2}+
\Phi^*\Phi\right) \right. \cr \cr &&\left.
-\frac\lambda{4N}\left(\frac{\Psi^2}{2}+\Phi^*\Phi\right)^2
-\frac{1}{4} F_{\mu \nu} \;  F^{\mu \nu} \right]
\end{eqnarray}
\noindent with
\begin{eqnarray} \label{masconf}
&&M^2(\eta) = -\mu^2 C^2(\eta)- \frac{C''(\eta)}{C(\eta)} \quad ,
\quad D_{\mu} = \partial_{\mu}-ie A_{\mu} ~~; ~~ F_{\mu \nu} =
\partial_{\mu }A_{\nu}-\partial_{\nu}A_{\mu}\; ,
\end{eqnarray}
\noindent and the primes refer to derivatives with respect to
conformal time.  Obviously the conformal rescaling of the metric and
fields turned the action into that of a charged scalar field
interacting with a gauge field in \emph{flat Minkowski space-time},
but the scalar field acquires a time dependent mass
term\footnote{Here we neglect the effect of the conformal
anomaly\cite{dolgovanomaly}}. In particular, in the absence of
electromagnetic coupling, the equations of motion for the gauge
field $ A_{\mu} $ are those of a free field in flat space time. This
is the statement that gauge fields are \emph{conformally} coupled to
gravity and no generation of electromagnetic fields can occur from
gravitational expansion alone without coupling to other fields or
breaking the conformal invariance of the gauge sector. The
generation of electromagnetic fields must arise from a coupling to
other fields that are not conformally coupled to gravity, or by
adding extra terms in the Lagrangian that would break the conformal
invariance of the gauge fields\cite{turnerwidrow}.

The dynamics is determined by the Heisenberg equations of motion of
the neutral field $ \Psi$ and the charged fields $ \Phi
$~\cite{cosmo,scaling,magfiI}.  We will consider that at the onset
of the radiation dominated era, the system is in the symmetric high
temperature phase in local thermal equilibrium with a vanishing
expectation value for the scalar fields. In the absence of explicit
symmetry breaking perturbations the expectation value of the scalar
field will remain zero throughout the evolution, thus $\varphi\equiv
0$.

It is convenient to introduce the mode expansion of the charged
fields
\begin{equation}
\Phi_r(\eta ,\vec x)= \int \frac{d^3 k}{\sqrt{2 \, (2\pi)^3}}
 \left[ a_r(\vec k) \; f_k(\eta )\;
e^{i\vec k\cdot \vec x}+ b_r^\dagger(\vec k)\; f^*_k(\eta )\;
e^{-i\vec k\cdot \vec x} \right]\quad  , \quad r=1,\ldots,N \; .
\label{phidecompo}
\end{equation}
In leading order in the large $N$ limit, the Heisenberg equations of
motion for the charged fields translate into the following equations
of motion for the mode functions for
$\eta>\eta_R$~\cite{cosmo,scaling,magfiI}
 \begin{equation}
\left[\frac{d^2}{d\eta
^2}+k^2-M^2(\eta)+\frac{\lambda}{2}\varphi^2(\eta )+
\frac{\lambda}{2N}\langle\Phi^{\dagger}\Phi\rangle\right ] \;
f_k(\eta )=0 \;. \label{ecmod}
\end{equation}
With our choice of the initial state we find the backreaction term
to be given by\cite{magfiI}
\begin{equation}\label{backreaction}
\lambda\Sigma(\eta )\equiv\frac{\lambda}{2N}\langle\Phi^{\dagger}
\Phi\rangle = \frac{\lambda}{4}\int \frac{d^3k}{(2\pi)^3}\;|f_k(\eta
)|^2[1+2n_k]\;.
\end{equation}
This expectation value features quadratic and logarithmic UV
divergences which are absorbed in the mass and  coupling
renormalization\cite{scaling}.

After renormalization and in terms of dimensionless quantities, the
non-equilibrium dynamics of the charged scalar fields is determined
by \cite{cosmo,scaling,magfiI},
\begin{eqnarray}\label{ecmov}
&&\left[\frac{d^2}{d\eta ^2}+
\mathcal{M}^2(\eta)+q^2+\lambda\Sigma(\eta ) \right]f_q(\eta
)=0~~;~~ \label{modeeqnew}
\end{eqnarray}
\noindent with the effective time dependent mass given by
\begin{eqnarray}\mathcal{M}^2(\eta)&=&
C^2(\eta) \; \mu^2\left[\frac{T^2_R}{C^2(\eta) \;
T^2_c}-1\right]\label{mossoft}\\
T^2_c &=& \frac{24 \; \mu^2}{\lambda+3e^2} .\label{crittemp}
\end{eqnarray}
We see that $ \mathcal{M}^2(\eta) > 0 $ for early times when $
C(\eta) < T/T_c $. For later times $ C(\eta) > T/T_c $ and $
\mathcal{M}^2(\eta) $ becomes negative triggering the phase
transition at a time $\eta_c$ when $ C(\eta_c)=T/T_c$ since we can
neglect the nonlinear contribution $ \lambda\Sigma(\eta ) $ (recall
that $ \lambda \ll 1 $).

When $ \mathcal{M}^2(\eta) $ is negative, eq.(\ref{ecmod}) tells us
that the modes $ f_q(\eta) $ grow exponentially as\cite{magfiI}
\begin{equation}\label{asimod}
f_q(\eta )\buildrel{\tilde{\mu} \; \eta \gg 1}\over = a_q \;
e^{\frac12( \tilde{\mu} \; \eta)^2} \; \left(\tilde{\mu} \;
\eta\right)^{ -\frac{q^2}{\tilde{\mu}^2}-1} \left[ 1 + {\cal
O}\left(\frac{1}{ \tilde{\mu}^2 \; \eta^2}\right) \right]
\end{equation}
where $\tilde\mu \equiv \sqrt{\mu \;  H_R} $ and $H_R$  is the
Hubble constant at the reheating time, $H_R=\eta_R^{-1}$.

This growth continues till the nonlinear term in eq.(\ref{ecmov}) $
\Sigma(\eta ) = \frac1{2N}\langle\Phi^{\dagger} \Phi\rangle $ cannot
be neglected anymore and stops the unstabilities. An scale invariant
stage follows\cite{scaling}. In this scaling stage the tree level
mass term $ - C^2(\eta) \; \mu^2 $ is compensated by the
backreaction of the quantum fluctuations as follows\cite{scaling},
\begin{equation}\label{conscond}
\lambda\Sigma(\eta)-\mu^2
C^2(\eta)\stackrel{\eta\rightarrow\infty}{=}-\frac{15}{4\eta^2} \; .
\end{equation}
The mode functions can be written during the scaling stage in terms
of Bessel functions
\begin{equation}\label{funbes}
f_k(\eta)=A_k\; \eta^{5/2}\; \frac{J_2(x)}{x^2}+B_k\; \frac{x^2
N_2(x)}{\eta^{3/2}}  \; ,
\end{equation}
where $ x=k\; \eta $ is the scaling variable. The correlation length
of the scalar field is of the order of the Hubble radius in this
stage.

\subsection{Gauge field dynamics}

The electric conductivity is very large in the high temperature
plasma formed after the second order phase transition and dominates
the dynamics of the gauge field. The conductivity is obtained from
the imaginary part of the photon polarization and it is dominated by
charged particles of momenta $ p\sim T $ in the loop with exchange
of photons of momenta $eT < k \ll T$. The effect of Debye (electric)
and dynamical (magnetic) screening via Landau damping is crucial
leading to the expression\cite{condu},
\begin{equation}\label{sigmacond}
\sigma(\eta) = \frac{\sigma_R}{C(\eta)} \quad , \quad \sigma_R =
\frac{\mathcal{C} \; N \;  T}{\alpha\ln\frac{1}{\alpha \; N}}
\end{equation}
where $ \mathcal{C} = 15.698\ldots $. Such large conductivity leads
to dissipative processes which severely hinders magnetogenesis (see
\cite{magfiI}) and also introduces the diffusion length scale which
could limit the correlation of the magnetic fields that are
generated.

As a consequence, we found for the photon causal correlator for $ k
\ll \sigma_R $
\begin{equation}
\mathcal{D}_R(\eta,\eta';k) = \theta(\eta-\eta') \;
\frac{e^{-\frac{k^2}{\sigma_R}(\eta-\eta')}}{\sigma_R} \;
.\label{DC}
\end{equation}

\section{Magnetic field spectrum }

The magnetic energy at wavenumber $ k $ is given by the symmetric
equal time limit
\begin{equation}\label{SB}
S_B(\eta,k)=\frac12\lim_{\eta'\to \eta }\int d^3x <\{\hat B^i(\eta
,\vec x), \hat B^i(\eta ',\vec 0)\}>_\rho e^{i\vec k\cdot \vec x}\;,
\end{equation}
where $\{\;,\;\}$ denotes the anti commutator,  $B(\eta ,\vec x)$
above is a \emph{Heisenberg operator} and the expectation value is
in the initial density matrix.

The \emph{physical} magnetic energy density  stored on
\emph{comoving} length scales larger than a given $L$ is given by
\begin{equation}
\Delta\rho_B(L,\eta )= \frac{1}{2\pi^2 }\int_0^{\frac{2\pi}{L}} k^2
\; S_B(\eta ,k) \; dk   \; .
\end{equation}
$\Delta \rho_B(L,\eta)$ stands for the contribution from the
non-equilibrium generation (subtracting the local thermodynamic
equilibrium contribution), a quantity of cosmological relevance to
assess the relative strength of the generated magnetic field is
given by the ratio of the power on scales larger than $L$ to the
energy density in the radiation background
\begin{equation}\label{ratio}
r(L,\eta)= \frac{\Delta\rho_B(L,\eta)}{\rho_{\gamma}(\eta)} \quad  ,
 \quad \rho_\gamma =\frac{\pi^2 T^4_R}{15 } \; .
\end{equation}
The explicit field theoretic evaluation of $ S_B(\eta,k) $ is given
in ref.\cite{magfiI}. The leading contribution to the power spectrum
generated by non-equilibrium fluctuations results expressed in terms
of the mode functions $ f_k(\eta) $ as follows,
\begin{eqnarray}\label{SButil}
&&S_B(\eta,k)=(1+2n_0)^2 \;\frac{\alpha N~k^2}{\pi ~\sigma^2_R} \;
e^{-\frac{2k^2}{\sigma_R}\eta} \; \int_0^{\infty} q^4 dq~ d(\cos
\theta) \; (1-\cos^2\theta) \cr \cr &&
\left|\int_{\eta_R}^{\eta}e^{\frac{k^2}{\sigma_R}\eta_1}\;
f_q(\eta_1)\; f_{|\vec q+\vec k|}(\eta_1)\; d\eta_1\right|^2 \; .
\end{eqnarray}
where $ \theta $ is the angle between the vectors $ \vec q $ and $
\vec k $ and we have used the result that both the spinodal stage as
well as the scaling stage is dominated by the long-wavelength  modes
that acquire non-perturbatively large amplitudes\cite{magfiI}.

We then compute $ r(L,\eta) $ using the scaling solution
eq.(\ref{funbes}) for the mode functions. The result can be recasted
for $ L \gg \eta  $ as\cite{magfiI},
\begin{equation}\label{rreges}
r(\eta,L) = 3.665 \times 10^4 ~
\left[\frac{\alpha}{\sigma_0}\right]^3 \; \frac1{[L \, T_R]^5} \;
\left[\frac{\mu}{\sqrt{\lambda} \; T(\eta)}\right]^4 \;
\left[\frac{M_*}{T(\eta)}\right]^3 ~.
\end{equation}
where
$$
T(\eta) = \frac{T_R}{C(\eta)} \quad ,  \quad M_* \simeq
\frac1{\sqrt{G}}  \simeq M_{Pl} \quad ,  \quad \sigma_0 =
\frac{15.698\ldots}{\ln \frac1{\alpha}} \; .
$$
Several important features of the above result are noteworthy:
\begin{itemize}
\item{{\bf i:} A typical nonperturbative behaviour $ \frac1{\lambda^2} $
arising from the phase transition.}
\item{{\bf ii:} $ r(\eta,L) $ grows with the symmetry breaking scale $ \mu $
as $ \mu^4 $ since the higher is $ \mu $ the longer is the scaling
stage.}
\item{{\bf iii:} The presence of the huge suppression factor $ [L \, T_R]^{-5}$.}
\end{itemize}

For the symmetry breaking scale $\mu \sim 10^{13}~\mbox{Gev}$ and
$\lambda \sim \alpha \sim 10^{-2}$, corresponding to a critical
temperature of order of  a GUT scale $T_c\sim 10^{15} \mbox{GeV}$
and assuming that the scaling regime lasts until the electroweak
phase transition scale, i.e. $\eta$ is such that $T(\eta)=T_{EW}\sim
10^2~\mbox{GeV}$. Then the factor
$$
\left(\frac{\mu}{\sqrt{\lambda}T_{EW}}\right)^4
\left(\frac{M_*}{T_{EW}}\right)^3\sim 10^{100}
$$
compensates for the factor $ [LT_R]^{-5} $. Taking $N$ and $g_*$ of
the order of $10$ we obtain
\begin{equation}\label{r-EW}
r[T(\eta),L] \simeq 10^{-34} \left(
\frac{L}{1~\mbox{Mpc}}\right)^{-5} \left[\frac{T_{EW}}{T(\eta)}
\right]^7 \; .
\end{equation}
Therefore,
 \begin{equation}\label{finrat}
r(T(\eta),L) \sim \left\{
\begin{array}{l}
  10^{-34}~\mbox{at\,the\,EW\,transition} \\
  10^{-14}~\mbox{at\,the\,QCD\,transition}\; .
\end{array}\right.
\end{equation}
Thus, the amplitude of the large scale magnetic fields turns to be
within the range necessary to be amplified by
 dynamo models.

It must be noticed that eq.(\ref{rreges})-(\ref{finrat}) provide the
long wavelength contributions to the magnetic field production which
dominate for $ L > 0.1$pc.

The results presented here (and in ref.\cite{magfiI}) arise from
quantum loop effects taking into account quantum processes that only
occur out of equilibrium. These  quantum processes are classically
forbidden. That is, they are virtual processes that take place
off-shell.

In summary, the large scale primordial magnetic fields generated
during the scaling stage after a phase transition are a plausible
mechanism to generate primordial magnetic fields which are further
amplified by the collapse of protogalaxies and by astrophysical
dynamos.

\begin{theacknowledgments}
 D.B.\ thanks the US NSF for support under
grant PHY-0242134,  and the Observatoire de Paris and LERMA for
hospitality. H. J. de V. thanks the Department of
Physics and Astronomy at the University of Pittsburgh for their
hospitality.
\end{theacknowledgments}

\end{document}